\documentstyle[epsfig]{article}

\begin{document}

\title{\LARGE \bf Shell model description of normal parity bands in
odd-mass heavy deformed nuclei}

\author{
C. Vargas$^1$\thanks{Electronic address: cvargas@fis.cinvestav.mx},
J. G. Hirsch$^2$\thanks{Electronic address: hirsch@nuclecu.unam.mx},
T. Beuschel$^3$\thanks{Electronic address: bthomas@rouge.phys.lsu.edu},
J. P. Draayer$^3$\thanks{Electronic address:
draayer@rouge.phys.lsu.edu},\\
{\small $^1$  Departamento de F\'{\i}sica, Centro de Investigaci\'on y de
Estudios Avanzados del IPN,}\\
{\small Apartado Postal 14-740 M\'exico 07000 DF, M\'exico}\\
{\small $^2$  Instituto de Ciencias Nucleares, Universidad Nacional
Aut\'onoma de M\'exico,}\\
{\small Apartado Postal 70-543 M\'exico 04510 DF, M\'exico }\\
{\small $^3$  Departament of Physics and Astronomy, Louisiana State
University,}\\
{\small Baton Rouge, LA 70803-4001, USA }
}

\date{\today}


\maketitle

\begin{abstract}
The low-energy spectra and B(E2) electromagnetic transition strengths of
$^{159}$Eu, $^{159}$Tb and $^{159}$Dy are described using the pseudo
SU(3) model.
Normal parity bands are built as linear combinations of SU(3) states,
which are the direct product of SU(3) proton and neutron states with
pseudo spin zero (for even number of nucleons) and pseudo spin 1/2 (for
odd number of nucleons). Each of the many-particle states have a
well-defined particle number and total angular momentum.
The Hamiltonian includes spherical Nilsson single-par\-ticle energies, the
quadrupole-quadrupole and pairing interactions, as well as three rotor
terms which are diagonal in the SU(3) basis.
The pseudo SU(3) model is shown to be a powerful tool to describe
odd-mass heavy deformed nuclei.

\bigskip
\noindent
{PACS numbers: 21.60.Fw, 21.60.Cs, 27.70.+q}
\end{abstract}

\vskip2pc

The shell model is a fundamental theory that is applicable in nuclear, atomic
and non-relativistic quark physics \cite{Vall91}. In its simplest formulation
it provides a natural explanation of magic numbers as shell closures and
the energy spectra of closed shell $\pm 1$ odd-mass nuclei \cite{May49,Hax49}.
Powerful computers and special algorithms for diagonalizing large matrices has
allowed systematic studies of nuclei of the {\em sd}-shell \cite{Bro88} and
{\em pf}-shell up to $A = 56$ \cite{Cau94}. New methods for solving large
scale shell-model problems in medium mass nuclei have also been developed
\cite{Hon96}. A shell-model description of heavy nuclei requires further
assumptions that include a systematic and proper truncation of the model space
\cite{Vall91}.

In light deformed nuclei the dominance of the quadrupole-quadrupole
interaction led to the introduction of the SU(3) shell model \cite{Ell58},
and with it a very natural means to truncate large model spaces.  Although
realistic interactions mix different irreducible representations (irreps), the
ground state wave function of well-deformed light nuclei normally consists of
only a few SU(3) irreps \cite{Aki69,Ret90,Tro96,Var98}. The strong spin-orbit
interaction renders the usual SU(3) scheme useless in heavy nuclei, but at the
same time pseudo-spin emerges as a good symmetry \cite{Hech69,Ari69,Rat73}.

Pseudo-spin symmetry refers to the experimental fact that single-particle
orbitals with $j = l - 1/2$ and $j = (l-2) + 1/2$ in the shell $\eta$ lie very
close in energy and can therefore be labeled as pseudo spin doublets with
quantum numbers $\tilde j = j$, $\tilde\eta = \eta -1$ and $\tilde l = l - 1$.
The origin of this symmetry has been traced back to the relativistic Dirac
equation \cite{Blo95,Gin97,Men98}. The pseudo SU(3) model capitalizes on the
existence of pseudo-spin symmetry.

In the simplest version of the pseudo SU(3) model, the intruder level with
opposite parity in each major shell is removed from active consideration and
pseudo-orbital and pseudo-spin angular momentum are assigned to the remaining
single-particle states. The coupling of a deformed rigid-rotor core with one
extra particle in a pseudo SU(3) orbital has been used to describe rotational
bands and electromagnetic properties of heavy odd-mass nuclei \cite{War90} and
identical normal and superdeformed bands \cite{Kre94}.

A fully microscopic description of low-energy bands in even-even nuclei has
been developed using the pseudo SU(3) model. The first applications used the
pseudo SU(3) as a dynamical symmetry, with a single SU(3) irrep describing the
whole yrast band up to backbending \cite{Dra82}. A comparison of quantum
rotor and microscopic SU(3) states \cite{Cas88} provided a classification
of the SU(3) irreps in terms of their transformation properties under $\pi$
rotations in the intrinsic frame \cite{Les87} and led to the construction of
a $K^2$ operator which plays a crucial role in the description of the gamma
band \cite{Naq90}.

On the computational side, the development of a computer code to calculate
reduced matrix elements of physical operators between different SU(3) irreps
\cite{Bah94} represented a breakthrough in the development of the pseudo SU(3)
model. For example, with this code it is possible to include pairing, which is
an SU(3) symmetry breaking interaction, in the Hamiltonian and exhibit its
close relationship with triaxiality \cite{Tro95,Bah95}. Full-space calculations
in the {\em pf}-shell \cite{Zuk95} in an SU(3) basis \cite{Var98} show that
for a description of the low-energy spectra of deformed nuclei the Hilbert
space can be truncated to leading irreps of the quadrupole-quadrupole and
spin-orbit (or pseudo spin-orbit) interactions. However, the inclusion of a
pairing-type interaction is essential for a correct description of moments of
inertia in such a truncated space.

Once a basic understanding of this overall structure was achieved, a powerful
shell-model theory for a description of normal parity states in heavy deformed
nuclei emerged. For example, the low-energy spectra of many Gd and Dy isotopes,
their B(E2) and  B(M1) transitions strengths for both their scissors and twist
modes \cite{Rom98} and their fragmentation were successfully described with a
realistic Hamiltonian \cite{Beu98}.

In the present letter we introduce a refined version of the pseudo SU(3)
formalism which uses a realistic Hamiltonian with single-particle energies
plus quadrupole-quadrupole and monopole pairing inteactions with strengths
taken from known systematics. The model is applied to three odd-mass rare earth
nuclei: $^{159}$Eu, $^{159}$Tb and $^{159}$Dy. The results represent a full
implementation of the very ambitious program implied in first applications of
the pseudo SU(3) model to odd-mass nuclei performed nearly thirty years ago
\cite{Rat73}.

Many-particle states of $n_\alpha$ active nucleons in a given normal
parity shell $\eta_\alpha$, $\alpha = \nu$ or $\pi$, can be classified by
the following chains of groups:

\begin{eqnarray}
~ \{ 1^{n^{N}_\alpha} \} ~~~~~~~ \{ \tilde{f}_\alpha \} ~~~~~~~\{ f_\alpha
\} ~\gamma_\alpha ~ (\lambda_\alpha , \mu_\alpha ) ~~~ \tilde{S}_\alpha
~~ K_\alpha  \nonumber \\
U(\Omega^N_\alpha ) \supset U(\Omega^N_\alpha / 2 ) \times U(2) \supset
SU(3) \times SU(2) \supset \nonumber \\
\tilde{L}_\alpha  ~~~~~~~~~~~~~~~~~~~~~ J^N_\alpha ~~~~ \nonumber \\
SO(3) \times SU(2) \supset SU_J(2),
\label{eq:chains}
\end{eqnarray}

\noindent where above each group the quantum numbers that characterize its
irreps are given and $\gamma_\alpha$ and $K_\alpha$ are multiplicity
labels of the indicated reductions.

The most important configurations are those with highest spatial symmetry
\cite{Dra82,Var98}. This implies that $\tilde{S}_{\pi , \nu} = 0$ or
$1/2$, that is, only configurations with pseudo spin zero for even number of
nucleons and $1/2$ for odd number of nucleons are taken into account.

We will describe $^{159}$Tb as a first example. It has 15 protons and 12
neutrons in the 50-82 and 82-126 shells, respectively.  The number of nucleons
in normal ({\em N}) and abnormal ({\em A}) parity orbitals is determined by
filling the Nilsson levels with a pair of particles for $\beta \sim 0.25$ in
order of increasing energy. This gives

\begin{eqnarray}
n^N_\pi = 9, ~~~ n^A_\pi = 6, ~~~ n^N_\nu = 8, ~~~ n^A_\nu = 4
\end{eqnarray}

\noindent After decoupling the pseudospin in Eq. (\ref{eq:chains}) we get
$ \{ \tilde{f}_\pi \} = \{ 2^4 1 \}, \{ \tilde{f}_\nu \} = \{ 2^4 \} $
with $\tilde{S}_\pi = 1/2$ and $\tilde{S}_\nu = 0$. Table I lists the 15
pseudo
SU(3) irreps, with the largest value of the Casimir operator $C_2$, which were
used in this calculation.

\vskip .25cm
\noindent Table I
\vskip .25cm

The Hamiltonian contains spherical Nilsson single-particle terms for
protons and neutrons ($H_{sp,\pi[\nu]}$), the quadrupole-quadrupole ($\tilde
Q \cdot \tilde Q$) and pairing ($H_{pair,\pi[\nu]}$) interactions as well as
three `rotor-like' terms which are diagonal in the SU(3) basis:

\begin{eqnarray}
 H & = & H_{sp,\pi} + H_{sp,\nu} - \frac{1}{2}~ \chi~ \tilde Q \cdot
         \tilde Q - ~ G_\pi ~H_{pair,\pi} ~\label{eq:ham} \\
   &   & - ~G_\nu ~H_{pair,\nu} + ~a~ K_J^2~ +~ b~ J^2~ +~ A_{asym}~
         \tilde C_2 . \nonumber 
\end{eqnarray}

\noindent The term proportional to $K_J^2$ breaks the SU(3) degeneracy of the
different K bands \cite{Naq90}, the $J^2$ term represents a small correction
to fine tune the moment of inertia, and the last $\tilde C_2 $ term is
introduced to distinguish between SU(3) irreps with $\lambda$ and $\mu$  both
even from the others with one or both odd
\cite{Les87}.

The Nilsson single-particle energies as well as the pairing and
quadrupole-quadrupole interaction strengths were taken from systematics
\cite{Rin79,Duf96}; only $a$ and $b$ were used for fitting. Parameter
values are listed in Table II and are consistent with those used in the
description of neighboring even-even nuclei \cite{Beu98}.

\vskip .25cm
Table II
\vskip .25cm

Figure 1(a) shows the calculated and experimental \cite{NNDC} $K =
{\frac 3 2}, ~{\frac 5 2}$ and ${\frac 1 2}$ bands for $^{159}$Tb. The
agreement between theory and experiment is in general excellent.
The model predicts a continuation of the $K = {\frac 5 2}$ band and
over-emphasizes staggering in the  $K = {\frac 1 2}$ band.

\vskip .25cm
Figure 1
\vskip .25cm

The role played by each term in the Hamiltonian will be discussed in detail
elsewhere \cite{Var99}. In this letter we wish to emphasize that the pairing
interaction is absolutely essential despite the strong truncation of the
Hilbert space.  To this end we present in part (b) of Fig. 1 the low-energy
spectra of $^{159}$Tb with the same Hamiltonian  {\em except that the pairing
interaction has been turned off}. It clearly exhibits the importance of the
pairing interaction in building up the correct moment of inertia: the spectra
{\em without} pairing is strongly compressed. It can also be seen that pairing
affect the other energies in a similar way with an overall effect that
resembles the introduction of a multiplicative factor in the Hamiltonian. We
conclude that the proposed truncation scheme is justified and works as
expected.

Theoretical and experimental \cite{NNDC} B(E2) transition strengths between
yrast  states in $^{159}$Tb are shown in Table III. The E2 transition
operator
that was used is given by \cite{Dra82}

\begin{equation}
Q_\mu = e_\pi Q_\pi + e_\nu Q_\nu \approx
e_\pi {\frac {\eta_\pi +1} {\eta_\pi}} \tilde Q_\pi +
e_\nu {\frac {\eta_\nu +1} {\eta_\nu}} \tilde Q_\nu ,
\end{equation}
with effective charges $e_\pi = 2.3, ~e_\nu = 1.3$. These values
are very similar to those used in the pseudo SU(3) description  of
even-even nuclei \cite{Dra82,Beu98}. They are larger than those used in
standard calculations of B(E2) strengths \cite{Rin79} due to the passive
role assigned to nucleons in unique parity orbitals, whose contribution to
the quadrupole moments is parametrized in this way.

In Figure 2 (a) we present the low-lying energy spectra of $^{159}$Eu,
including the $K = {\frac 5 2}, {\frac 3 2} $ and ${\frac 1 2}$ bands
built with 7 protons in the normal parity subshell $\tilde{\eta} = 3$ and
8 neutrons in $\tilde{\eta} = 4$. There is a good agreement between the
experimental \cite{NNDC} and theoretical results. The model predicts a second
${\frac 7 2}^+$ state in the $K = {\frac 3 2}$ band which is missing in the
experimental spectra, as well as several other states in the excited bands.

\vskip .25cm
Figure 2
\vskip .25cm

\noindent It is interesting to notice that the ground state in $^{159}$Tb is
${\frac 3 2}^+$ while in $^{159}$Eu it is ${\frac 5 2}^+$.
Reproducing this effect is one of the successes of this theory; realistic
single-particle-energies are required to get this ordering correct.

The low energy spectra of $^{159}$Dy is presented in Fig. 2(b).
There are three bands, with $K = {\frac 3 2}, {\frac 5 2}$ and
${\frac 1 2}$, respectively. As in the other cases
the agreement between theory and experiment is remarkably good.
In the $K = {\frac 3 2}$ ground state band the ${\frac {17} 2}^-$ state is
predicted to have an energy higher than the experimentally observed one. This
departure of the experimental ground state band from the rigid rotor behavior
may be related with a band crossing. The possibility of describing it by
increasing the Hilbert space is under investigation. In the $K = {\frac 1 2}$
band the  ${\frac{3}{2}}^-$ state lies higher than its ${\frac{5} 2}^-$
partner which  contradicts the experimental results. As in the other cases,
the model predicts several excited levels that are as yet undetected.

It has been shown that normal parity bands in odd-mass heavy deformed nuclei
can be described quantitatively using the pseudo SU(3) model. Only a few
representations with largest $C_2$ values and pseudo spin 0 or 1/2 are
needed. The Hamiltonian uses Nilsson single-particle energies,
quadrupole-quadrupole and pairing interactions with strengths fixed by
systematics, and three small rotor terms which with the others yield
excellent results for energies and B(E2) values in A=159 nuclei.

This work exhibits the usefulness of the pseudo SU(3) model as a shell model,
one which can be used to describe deformed rare-earth and actinide isotopes by
performing a symmetry dictated truncation of the Hilbert space. It opens up
the possibility of a more detailed microscopic description of other properties
of heavy deformed nuclei, both with even and odd protons and neutrons numbers,
like g-factors, M1 transitions, and beta decays.

This work was supported in part by Conacyt (M\'exico) and the U.S. National
Science Foundation.

\newpage

{\bf Table Captions}\\

Table I: The 15 pseudo SU(3) irreps used in the description of $^{159}$Tb
bands.

Table II: Parameters used in Hamiltonian~\ref{eq:ham}.

Table III: Theoretical and experimental B(E2) transition strengths
for $^{159}$Tb.

\bigskip

{\bf Figure Captions} \\

Fig. 1: Energy spectra of $^{159}$Tb, `Exp' represents the experimental
results and `Theo' the calculated ones. Insert (a) shows the energies
obtained with the Hamiltonian parameters listed in Table II, insert (b)
shows the energies obtained without pairing.

Fig. 2: (a) Energy spectra of $^{159}$Eu and (b) $^{159}$Dy, with
the same convention of Fig. 1.

\newpage

\begin{table*}[h]
\begin{tabular}{cc|cccccc}
$(\lambda_\pi, \mu_\pi )$ &$(\lambda_\nu, \mu_\nu )$ &
\multicolumn{6}{c}{total $(\lambda, \mu )$ } \\ \hline
(10,4) &(18,4) & (28,8) &(29,6) &(30,4)& (31,2) & (32,0) & (26,9)\\
(11,2) &(18,4) & (29,6) & (30,4) & (31,2) \\
(10,4) &(20,0) & (30,4) \\ 
(11,2) &(20,0) & (31,2) \\
(7,7)  &(18,4) & (25,11) & (26,9) \\
(10,4) &(16,5) & (26,9) \\
(8,5)  &(18,4) & (26,9)
\label{eq:irreps}
\end{tabular}
~~~\\
\caption{}
\end{table*}

\begin{table*}[h]
\begin{tabular}{ccccccc}
           &${\chi}$&$G_\pi$&$G_\nu$&   $a$   &  $b$  &$A_{asym}$\\
\hline
$^{159}$Eu &0.00753 & 0.132 & 0.106 & -0.0508 & 0.0009& 0.0008   \\
$^{159}$Tb &0.00753 & 0.132 & 0.106 &  0.0198 &-0.0031& 0.0008   \\
$^{159}$Dy &0.00753 & 0.132 & 0.106 &  0.0048 & 0.0006& 0.0008
\end{tabular}
~~\\
\caption{}
\label{fig:table2}
\end{table*}

\begin{table*}[h]
\begin{tabular}{ccc}
$J^+ \rightarrow {(J+2)}^+$ & Theo $(e^2 b^2)$ & Exp $(e^2 b^2)$ \\ \hline
$\frac{3}{2}^+ \rightarrow \frac{7}{2}^+$  & 1.6503 & $1.4736 \pm 0.2047$ \\
$\frac{5}{2}^+ \rightarrow \frac{9}{2}^+$  & 2.0553 & $1.8590 \pm 0.1023$ \\
$\frac{7}{2}^+ \rightarrow \frac{11}{2}^+$ & 2.1966 & $2.2180 \pm 0.0537$ \\
$\frac{9}{2}^+ \rightarrow \frac{13}{2}^+$ & 2.2464 & $2.3280 \pm 0.0645$ \\
$\frac{11}{2}^+ \rightarrow \frac{15}{2}^+$& 2.2568 & $2.1080 \pm 0.1433$ \\
$\frac{13}{2}^+ \rightarrow \frac{17}{2}^+$& 1.4542 & $1.9867 \pm 0.1316$
\\ &&\\ \hline
$J^+ \rightarrow {(J+1)}^+$ & Theo $(e^2 b^2)$ & Exp $(e^2 b^2)$ \\ \hline
$\frac{3}{2}^+ \rightarrow \frac{5}{2}^+$   & 2.9988 & $2.8013 \pm 0.1458$ \\
$\frac{5}{2}^+ \rightarrow \frac{7}{2}^+$   & 1.6914 & $1.5691 \pm 0.3411$ \\
$\frac{7}{2}^+ \rightarrow \frac{9}{2}^+$   & 1.0471 & $0.7483 \pm 0.0831$ \\
$\frac{9}{2}^+ \rightarrow \frac{11}{2}^+$  & 0.7084 & $0.6877 \pm 0.0675$ \\
$\frac{11}{2}^+ \rightarrow \frac{13}{2}^+$ & 0.5201 & $0.3761 \pm 0.0477$ \\
$\frac{13}{2}^+ \rightarrow \frac{15}{2}^+$ & 0.3726 & $0.4386 \pm 0.0760$ \\
\end{tabular}
~~~\\
\caption{}
\label{fig:table3}
\end{table*}

\newpage

\begin{figure*}[h]
\epsfxsize=6.5in
\centerline{Fig. 1}
\centerline{\epsfbox{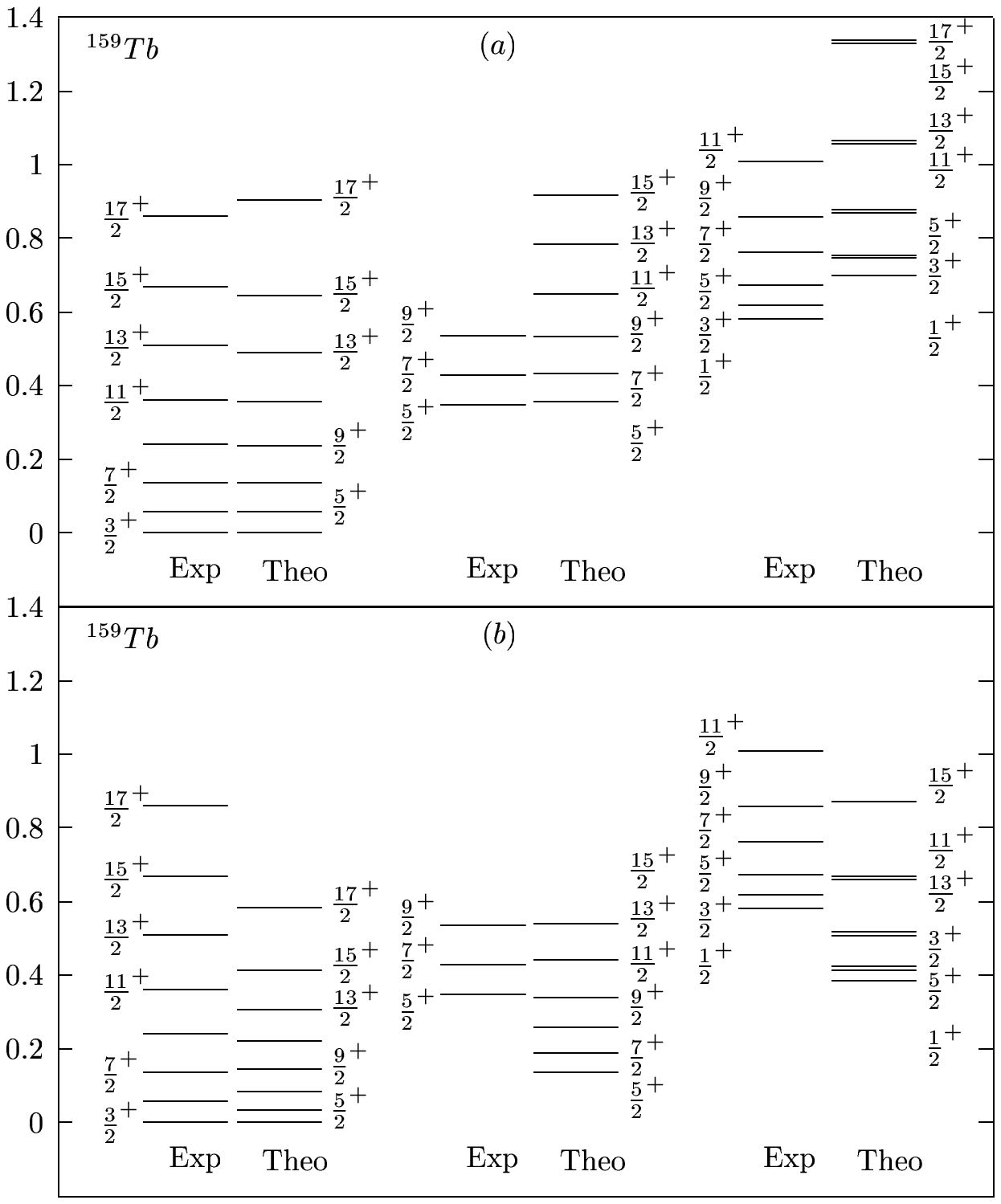}}
\end{figure*}

\begin{figure*}[h]
\centerline{Fig. 2}
\epsfxsize=6.5in
\centerline{\epsfbox{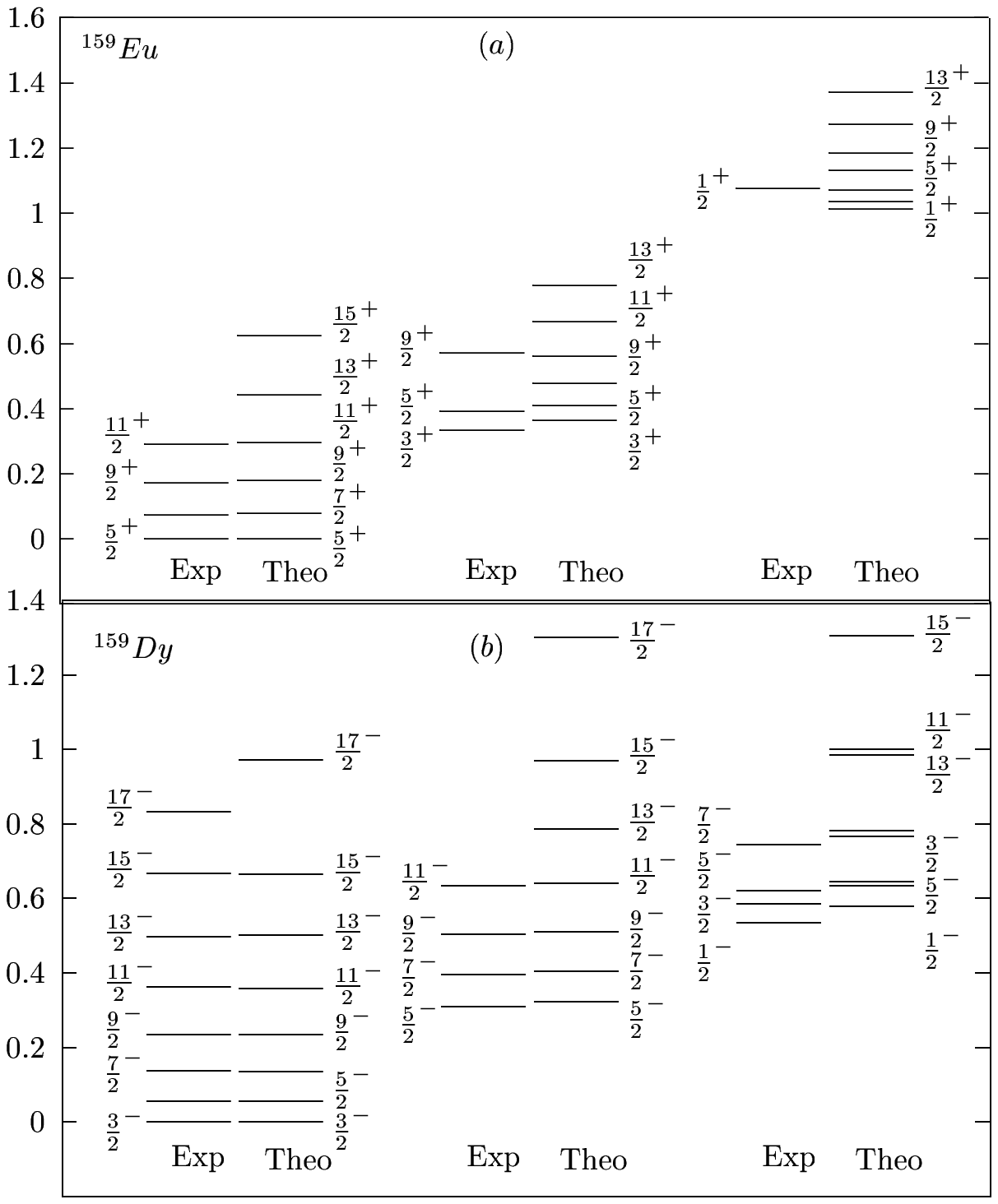}}
\end{figure*}

\end{document}